\documentclass{webofc}
\usepackage[varg]{txfonts}   % Web of Conferences font
\begin{document}
\title{The hydrostatic mass bias in {\sc The Three Hundred} clusters}
%
% subtitle is optionnal
%
%\subtitle{Is it affected by mergers?}

\author{\firstname{Giulia} \lastname{Gianfagna}\inst{1,2}\fnsep\thanks{\email{giulia.gianfagna@uniroma1.it}} \and 
	\firstname{Elena} \lastname{Rasia}\inst{3,4} 
	\and
	\firstname{Weiguang} \lastname{Cui}\inst{5}
\and
	\firstname{Marco} \lastname{De Petris}\inst{1}
\and
	\firstname{Gustavo} \lastname{Yepes}\inst{6}
}

\institute{Dipartimento di Fisica, Sapienza Università di Roma, Piazzale Aldo Moro 5, I-00185 Rome, Italy
\and
           INAF, Istituto di Astrofisica e Planetologia Spaziali, via Fosso del Cavaliere 100, 00133 Rome, Italy
\and
           IFPU - Institute for Fundamental Physics of the Universe, Via Beirut 2, 34014 Trieste, Italy 
\and
           INAF Osservatorio Astronomico di Trieste, via Tiepolo 11, I-34131, Trieste, Italy
\and
           Institute for Astronomy, University of Edinburgh, Royal Observatory, Edinburgh EH9 3HJ, UK
\and
           Departamento de F\'{\i}sica Te\'orica and CIAFF, M\'odulo 8, Facultad de Ciencias, Universidad Aut\'onoma de Madrid, 28049 Madrid, Spain
          }

\abstract{The assumption of Hydrostatic equilibrium (HE) is often used in observations to estimate galaxy clusters masses. We use a set of almost 300 simulated clusters from {\sc The Three Hundred} Project, to estimate the cluster HE mass and the bias deriving from it. We study the dependence of the bias on several dynamical state indicators across a redshift range from 0.07 to 1.3, finding no dependence between them. Moreover, we focus our attention on the evolution of the HE bias during the merger phase, where the bias even reaches negative values due to an overestimation of the mass with HE.
}
\maketitle
\section{Introduction}
\label{intro}

Galaxy clusters are the most massive, gravitationally bound structures in the Universe and they are a powerful tool in cosmology. In particular, their mass is fundamental for the estimation of cosmological parameters. In observations, one way to estimate this mass is through X-rays and SZ (Sunyaev-Zeldovich) effect, from which the temperature, density and pressure profiles of the hot gas between the galaxies (Intra Cluster Medium, ICM) are extracted, then the assumption of hydrostatic equilibrium (HE) has to be made in order to estimate their mass \cite{kravtsov_rev}. This method usually leads to an underestimation of the mass, as shown in numerical simulations \cite{gianfagna2021}. 
The lack of spherical symmetry and non-thermal motions in the gas lead to a violation of the HE assumption. These are usually present in the object outskirts, in major-merger events and in general in disturbed clusters \cite{rasia, pearce2019, ansarifard, contreras}. 
The presence of a correlation between hydrostatic mass bias and cluster dynamic state is still a matter of debate \cite{ rasia, ansarifard}, in this work we seek to understand this in detail by looking directly to the bias change along the evolution of  major-merger events.

\section{{\sc The Three Hundred} Project}
\label{sec-1}

The clusters in this work are the most massive clusters in {\sc The Three Hundred} Project \cite{Cui2018}, they come from a set of 324 Lagrangian regions from the dark-matter (DM) only simulation MultiDark \cite{Klypin2016}. We analyse 9 redshifts in the [0.07, 1.32] range.

{\sc The Three Hundred} set is a volume-limited mass-complete sample, the objects masses $M_{500}$\footnote{The mass with the subscript 500 is indicating the mass of a sphere with radius $R_{500}$ whose density is 500 times the critical density of the Universe $\rho_c = 3H(a)^2/8\pi G$, where $H(a)$ is the Hubble function.} are larger than $6.5\times10^{14} \rm M_{\odot}$. Initial conditions are generated at the redshift $z = 120$ by refining the mass resolution in the central region and degrading it in the outer part with multiple levels of mass refinement. The high-resolution dark-matter particle mass is equal to $m_{DM} = 1.27\times10^{9}h^{-1} \rm M_{\odot}$, while the initial gas mass is equal to $2.36\times10^8h^{-1} \rm M_{\odot}$. 

We consider the resimulated cluster sample with the hydrodynamical code GADGET-X \cite{Rasia2015}, which includes radiative processes, gas-cooling, star formation, supernovae thermal feedback, chemical evolution and enrichment, super massive black holes and their feedback. 

This simulation assumes a standard cosmological model according to the 2016 Planck results \cite{Planck2016}: $h = 0.6777$ for the reduced Hubble parameter, $n = 0.96$ for the primordial spectral index, $\sigma_8 = 0.8228$ for the amplitude of the mass density fluctuations in a sphere of $8 h^{-1}$ Mpc co-moving radius, $\Omega_{\Lambda} = 0.692885$, $\Omega_m = 0.307115$, and $\Omega_b = 0.048206$ respectively for the density parameters of dark energy, dark matter, and baryonic matter.

\section{Methods}
\label{sec-2}

In this section we show all the basics to estimate the hydrostatic masses and their bias with respect to the real mass.

\subsection{The Hydrostatic Equilibrium}
\label{subsec-3}

The hydrostatic equilibrium assumption assumes that the cluster is in equilibrium due to the balance between the gas thermal pressure and the gravitational force.
The HE mass, namely the total mass inside a sphere of radius $r$, can be written as
\begin{equation}
M_{\rm HE, SZ}(<r) = -\frac{r^2}{G \rho_{\rm g}(r)} \frac{\textrm{d} P_{\rm th}(r)}{ \textrm{d}r} ,
\label{eq:Mhe_P}
\end{equation}
it is based on the assumptions of spherical symmetry of the system and purely thermal gas pressure. In Eq.(\ref{eq:Mhe_P}) $G$ is the gravitational constant, $\rho_{\rm g}$ and $P_{\rm th}$ are the gas density and the thermal pressure of the gas. Assuming the equation of state of an ideal gas, it follows 
\begin{equation}
M_{ \rm HE, X}(<r) = - \frac{rk_{\rm B} T(r)}{G\mu m_p}\left[ \frac{\textrm{ d}\ln \rho_{\rm g} (r)}{\textrm{ d}\ln r} + \frac{\textrm{ d}\ln T(r)}{\textrm{ d}\ln r}\right].
\label{eq:Mhe_T}
\end{equation}
The real mass $M_{\rm true}$ of a simulated cluster can be easily computed by summing all the dark matter, stars and gas particle masses inside an aperture radius. The mass bias, at a specific radius, is defined as $b_{\rm SZ/X} = (M_{\rm true} - M_{\rm HE, SZ/X})/M_{\rm true}$. It can be either positive, when the HE mass is underestimating the cluster true mass, or negative, when there is an overestimation. A null bias suggests that the true mass is perfectly described by HE. 

\subsection{The ICM profiles}
\label{subsec-5}

The 3D radial profiles of pressure, temperature and gas density are estimated dividing the cluster in logarithmically equispaced radial bins, starting from the center (the place of the minimum potential well \cite{ansarifard}). The profiles are fitted in the radial range [0.2-3]$R_{500}$.

The 3D gas density profiles are estimated as the total gas mass in the spherical shell, divided by the shell volume. We use the Vikhlinin model \cite{vikh} to fit these profiles 
\begin{equation}
\rho_g(r) = \rho_0^2 \frac{(r/r_{\rm c})^{-a}}{ (1 + (r/r_{\rm c})^2)^{3b-a/2}} \frac{1}{ (1 + (r/r_{\rm s})^c)^{e/c} } + \frac{\rho_{02}^2}{ (1 + (r/r_{\rm c2})^{2})^{3b_2} },
\label{eq:rho_vikh}
\end{equation}
where $c$ is fixed at 3 and $e < 5$. The other 8 parameters are left free.

For the mass-weighted temperature profile we consider only particles with temperature $kT > 0.3 \rm keV$. The analytical model was introduced again by Vikhlinin in \cite{vikh}
\begin{equation}
T(r) = T_0 \times \frac{x + \tau}{x+1} \times \frac{(r/r_{\rm t})^{-a}}{ (1 + (r/r_{\rm t})^b)^{c/b} }
\label{eq:T_vikh}
\end{equation}
with $x = (r/r_{\rm cool})^{a_{\rm cool}}$. All the 8 parameters are free to vary.

To model the radial pressure profile we make use of the generalized Navarro-Frenk-White (gNFW) model \cite{nagai}
\begin{equation}
P(r) = \frac{P_0}{x^{c} (1 + x^{a})^{\frac{b - c}{a}}}
\label{eq:gNFW}
\end{equation}
with $x = r/r_s$ a dimensionless radial distance normalised to the scale radius $r_s = R_{500}/c_{500}$, where $c_{500}$ is the concentration parameter. This model has 5 free parameters.

The hydrostatic masses are estimated using the fitted profiles, instead of the numerical ones, to avoid the impact of fluctuations on the profiles (see \cite{gianfagna2021} for a deeper discussion). 

\subsection{Dynamical state}
\label{subsec-4}

We divide the clusters depending on their dynamical state thanks to two 3D estimators \cite{neto}: $f_s = M_{\rm sub}/M$, the fraction of cluster mass included in substructures; $\Delta_{\rm r} = |\textbf{r}_{\delta} - \textbf{r}_{\rm cm}| / R$, the offset between the central density peak, and the centre of mass of the cluster, normalized to the aperture radius. Each of the estimators is computed at the three nominal overdensities of 200, 500 and 2500. Both indicators are lower than 0.1 for relaxed cluster, larger than 0.1 for disturbed cluster and all the other cases are defined as hybrid clusters \cite{Cialone, DeLuca}. 

The information given separately by these indicators can be joint through the so called relaxation parameter $\chi_{\rm DS} = \sqrt{2 / [ ( \Delta_{\rm r}/0.1 )^2 + ( f_s/0.1 )^2 ]}$ \cite{DeLuca}. For a relaxed cluster $\chi_{DS}\geq 1$.

\section{Results}
\label{sec-6}

We study the HE bias dependencies on redshift, mass and dynamical state at the overdensities $R_{2500}$, $R_{500}$ and $R_{200}$.

\subsection{Bias correlations}
\label{subsec-7}

\begin{figure}[h]

    \includegraphics[width=.5\textwidth]{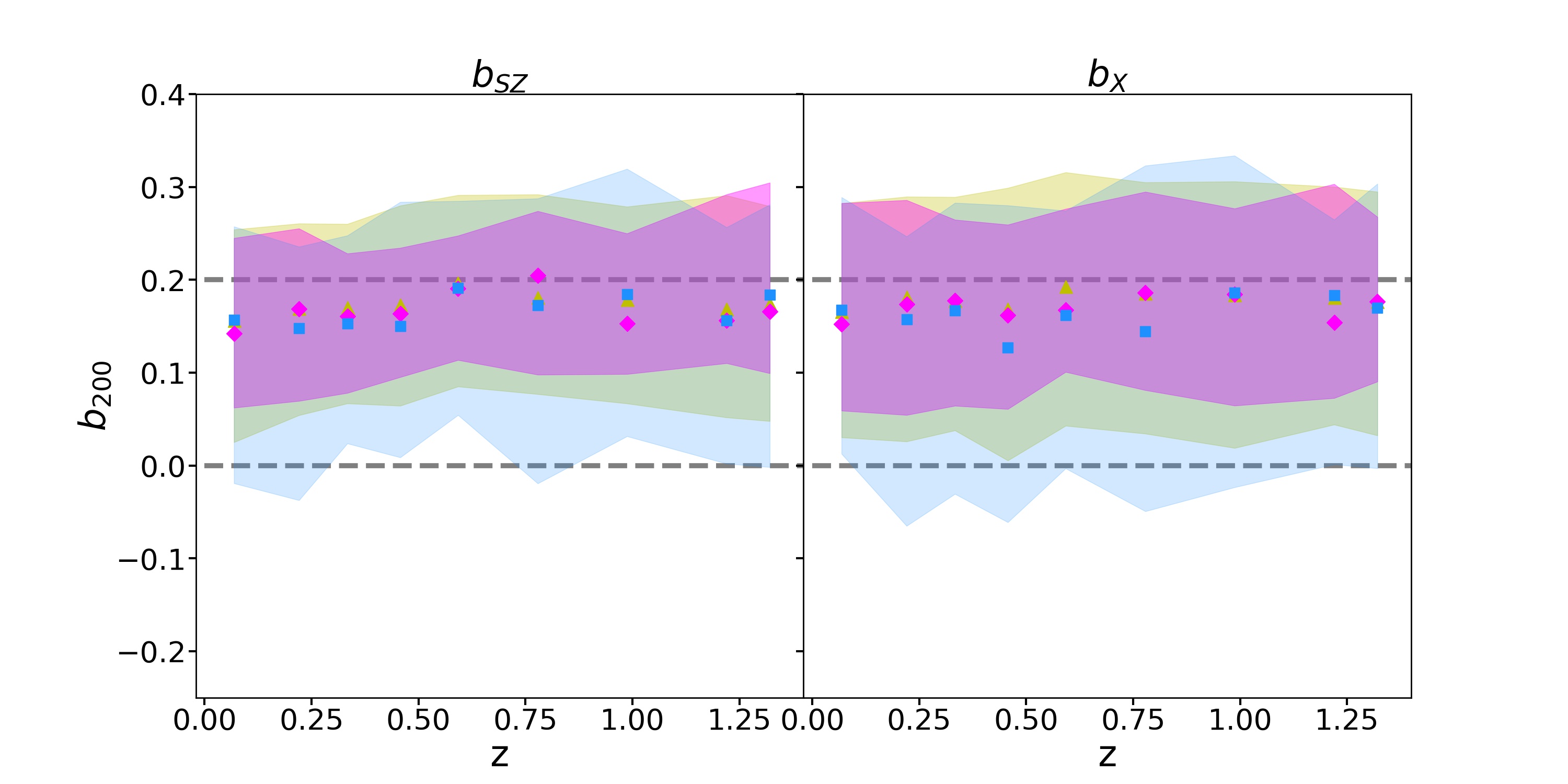}
    \includegraphics[width=.5\textwidth]{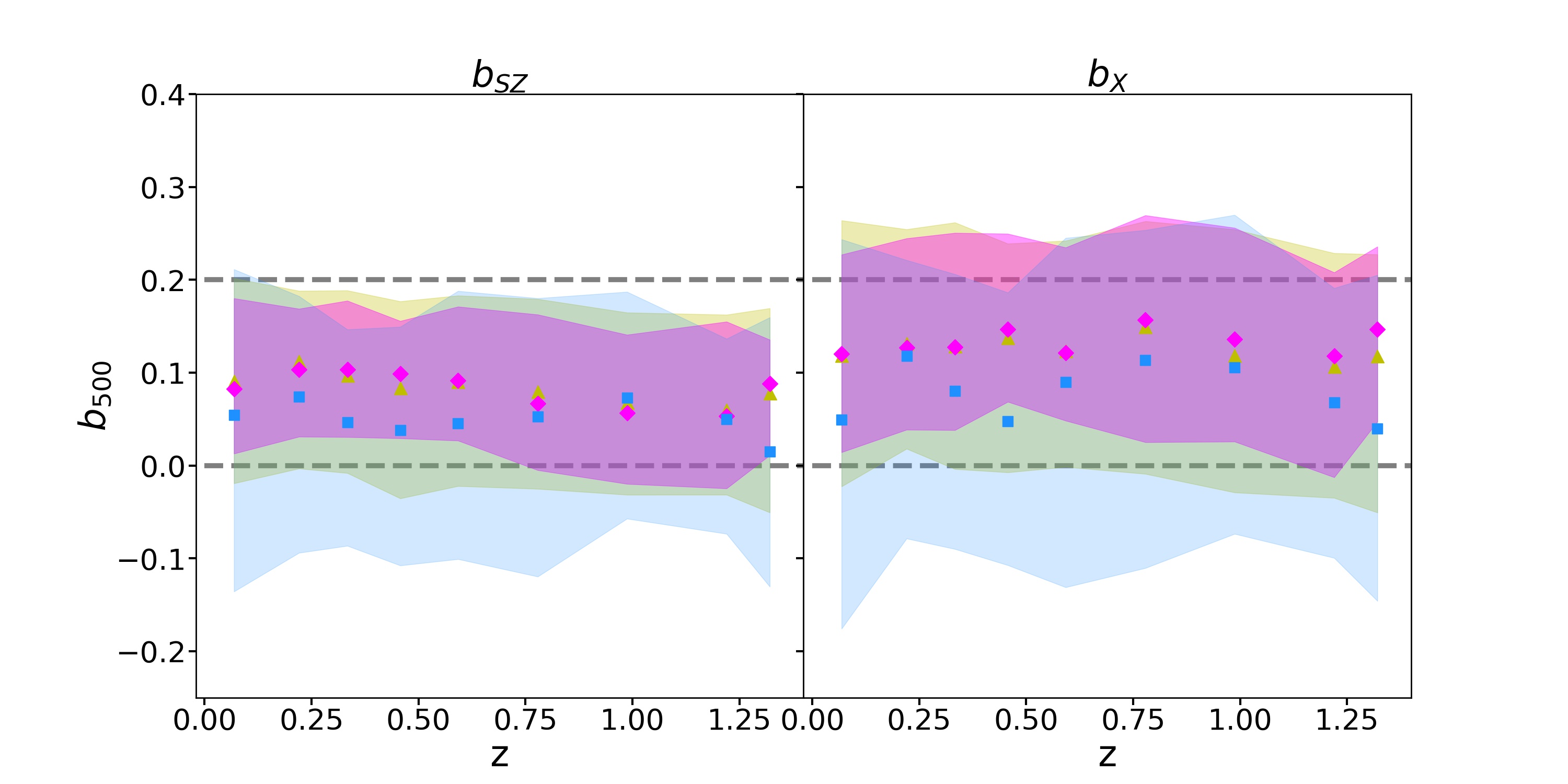} 

\begin{minipage}[c]{0.5\textwidth}
      \includegraphics[width=1\textwidth]{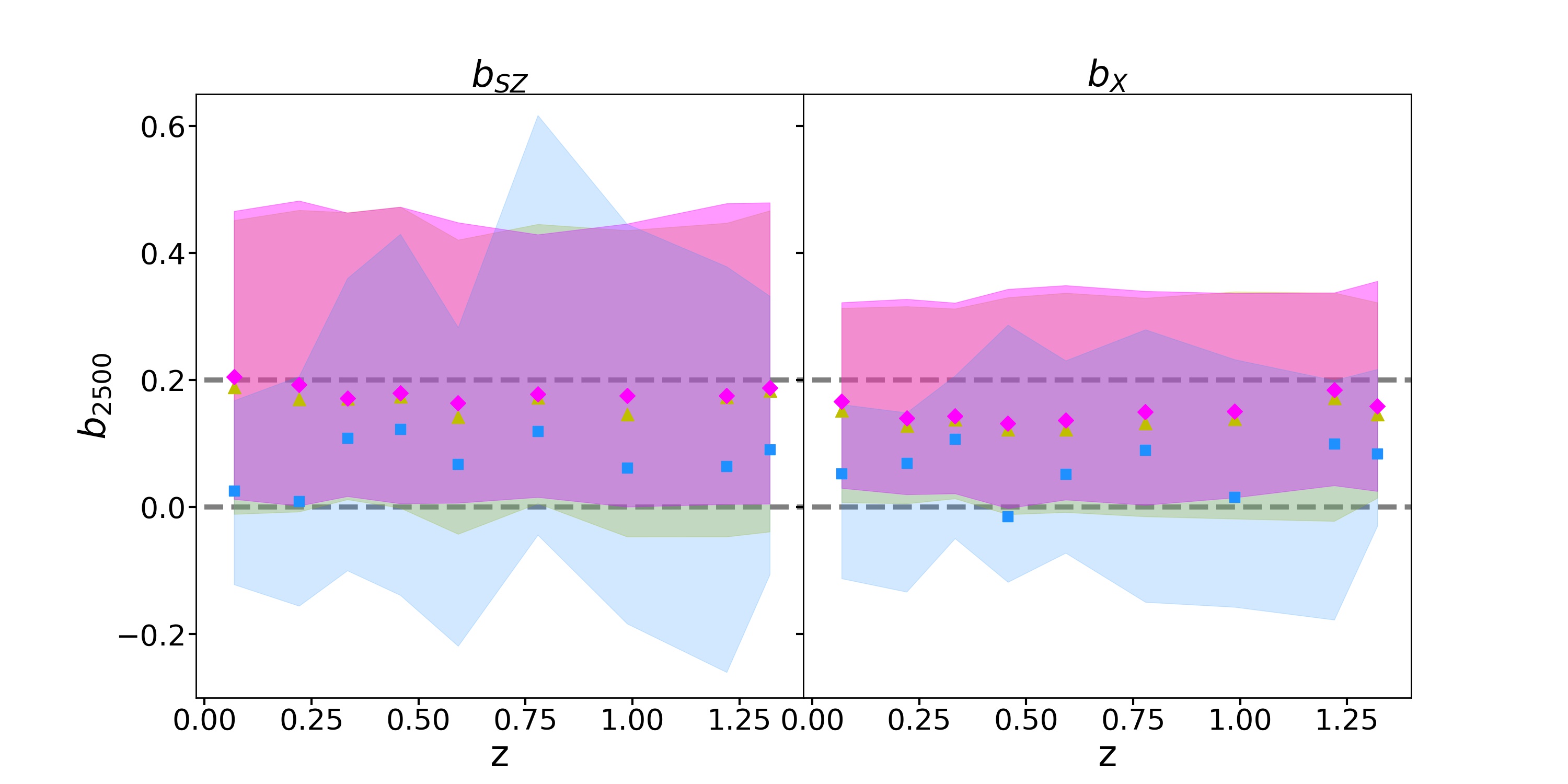} 
\end{minipage}
    \vspace{1cm}
\begin{minipage}[c]{0.45\textwidth}
    \caption{The redshift evolution of $b_ {\rm SZ}$ (left panels) and $b_{\rm X}$ (right panels). The median values of the bias for all clusters, relaxed and disturbed are represented with yellow triangles, magenta diamonds and blue squares respectively. The shaded regions represent the $16^{th}$ and $84^{th}$ percentiles. The bias estimated at $R_{200}$, $R_{500}$ and $R_{2500}$ are represented in the top left, top right and bottom panels respectively. The dashed lines represent 0 and 0.2 biases.} \label{fig-1}
  \end{minipage}

\end{figure}

The bias dependence on the redshift is presented in Fig. \ref{fig-1}. The disturbed clusters have the largest dispersion and have negative biases, independently from the overdensity radius. Near the cluster center, the deviations from HE are stronger than in the other overdensities \cite{pearce2019, gianfagna2021}, however the bias at $R_{2500}$ shows the largest scatter. 
No bias dependence on the redshift is evident inside the errors, in agreement with \cite{lebrun, salvati}.

\begin{figure}[h]
    \includegraphics[width=.5\textwidth]{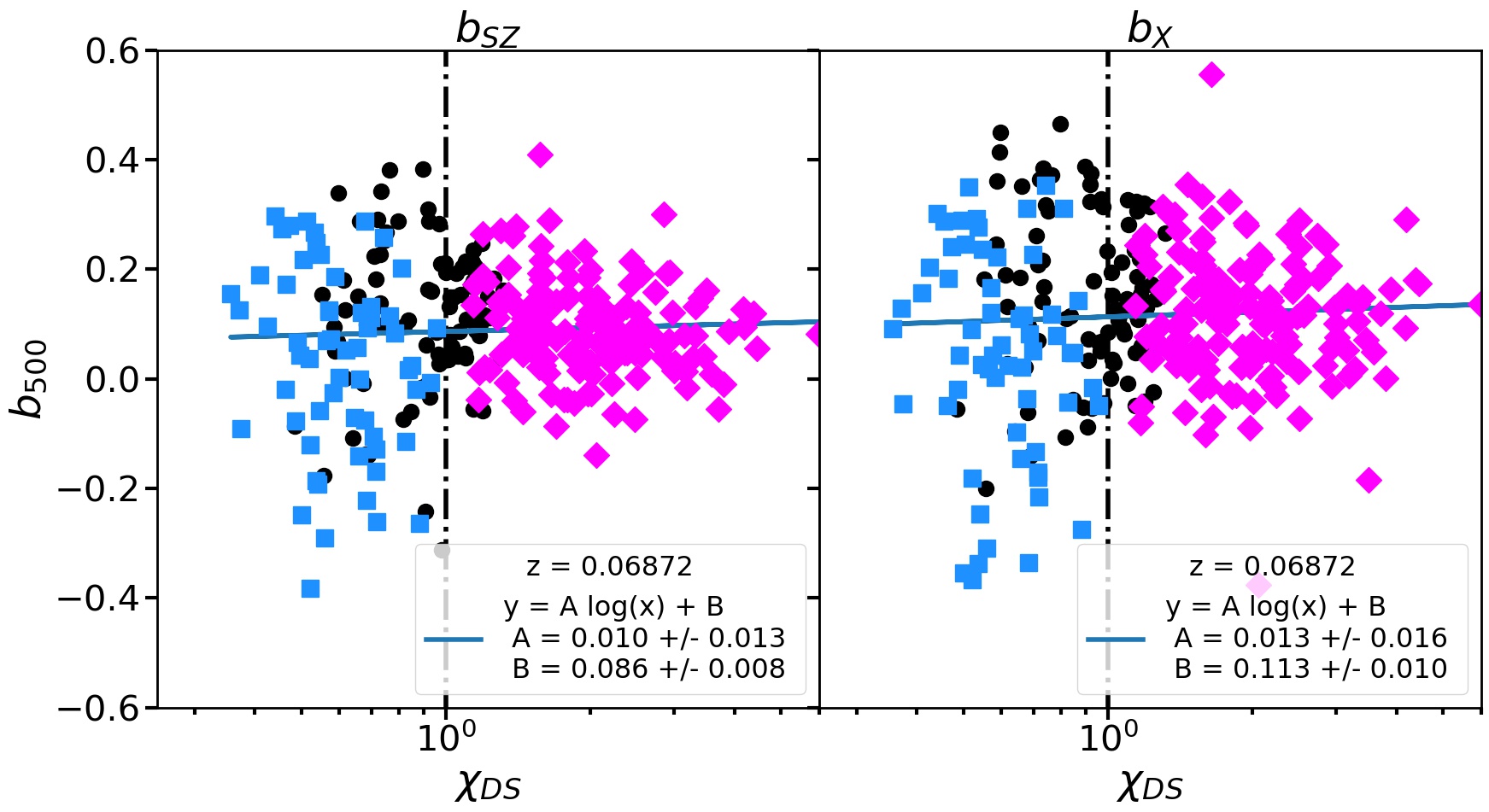}
    \includegraphics[width=.5\textwidth]{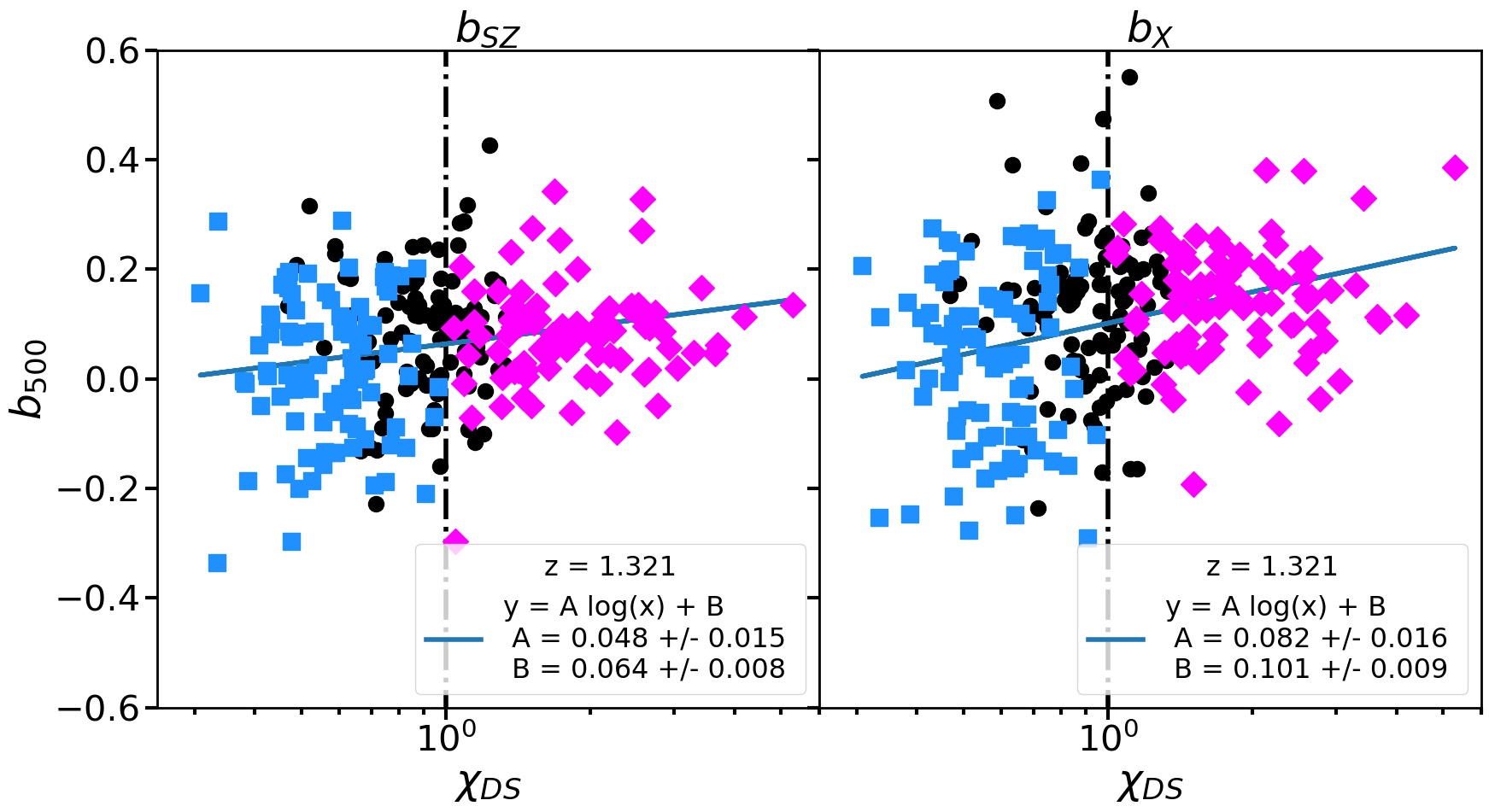}
    \caption{The biases are shown as a function of the relaxation parameter $\chi_{DS}$. The intermediate, relaxed and disturbed are represented with black dots, magenta diamonds and blue squares respectively. In the left panel we show the quantities at $z=0.07$ and in the right one at $z=1.3$.}
 \label{fig-2}
\end{figure}

The bias as a function of the relaxation parameter is represented in Fig.\ref{fig-2}, the logarithmic scale is used to inflate the data distribution in the small $\chi_{DS}$ range. The fit helps to spot any correlation, reducible to linear due to the narrow $\chi_{DS}$ scale. The Pearson correlation coefficients are 0.14 (p-value 0.02) for $b_{\rm SZ}$ and 0.23 (p-value 0.0001) for $b_{\rm X}$ at z=1.32; 0.01 (p-value 0.91) for $b_{\rm SZ}$ and 0.01 (p-value 0.81) for $b_{\rm X}$ at z=0.07. Fixing the confidence level at 0.05, we cannot say that the biases are correlated at z=0.07 (p-value $>0.05$), instead at $z=1.32$ there is indeed a mild correlation. At the other redshifts we have p-value $>0.05$, except for $z=0.46$, where the correlation coefficient points again to a mild correlation. However, the connection between the biases and the dynamical state can be affected by the use of the best fit model of the profiles, as this usually leads to a smoothing of any possible fluctuation. 
Also, the $b-\chi_{DS}$ dependency can be influenced by the thresholds chosen to determine the dynamical state of each cluster \cite{DeLuca}. Both these issues will most likely tend to reduce the 
potential correlation.
Moreover, also in this case, the disturbed clusters show a wider dispersion with respect to the relaxed ones \cite{rasia, ansarifard}.

\begin{figure}
  \begin{minipage}[c]{0.5\textwidth}
    \includegraphics[width=\textwidth]{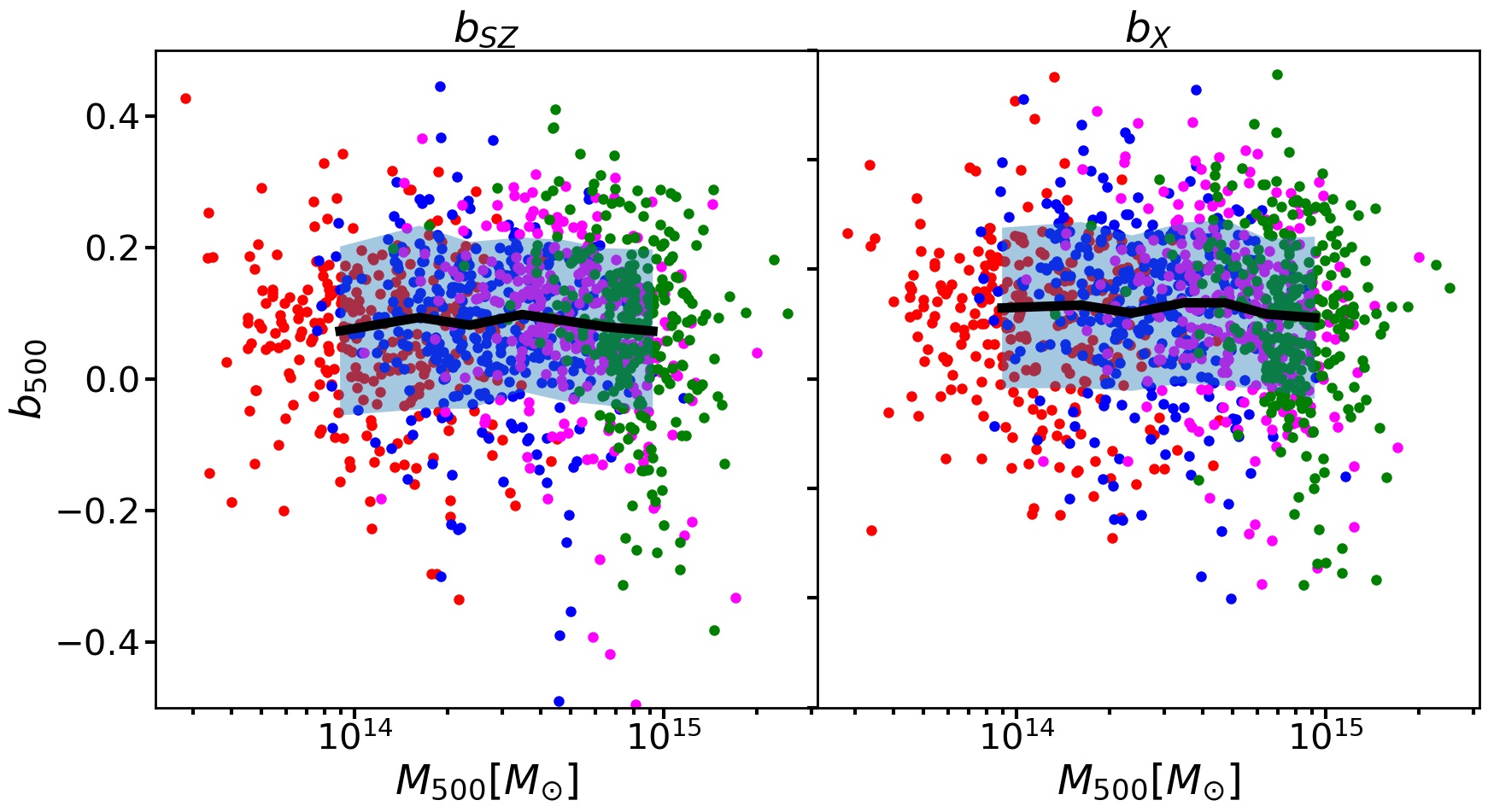}
  \end{minipage}\hfill
  \begin{minipage}[c]{0.42\textwidth}
    \caption{The biases are represented as a function of the cluster total mass at $R_{500}$. The redshifts 1.32, 0.78, 0.33, 0.07 are represented in red, blue, magenta and green respectively. The black line and the shaded region represent the binned median and 16th-84th percentiles.
    } \label{fig-3}
  \end{minipage}
\end{figure}

Finally, the bias dependence on the cluster mass at $R_{500}$ is represented in Fig.\ref{fig-3}. We show the biases at 4 redshifts (1.32, 0.78, 0.33, 0.07), and we find no dependence of the bias on the total mass of the clusters, as in \cite{lebrun, pearce2019, ansarifard, barnes2020}. The same happens also for $R_{2500}$ and $R_{200}$. However, the mass range considered is not wide enough to make definitive statements. 

\subsection{Hydrostatic bias and merging events}
\label{subsec-8}

Hydrodynamical simulations allow for tracking the entire history of a cluster, identifying the merging event and the moment it takes place. We define a merger as a very rapid increase in the halo mass. Using the fractional mass change $\Delta M / M = (M_f - M_i)/ M_i$, we consider major-merger, with $\Delta M / M \geq 1$, a merger that takes place in half of the dynamical time, with a 100\% mass increase \cite{contreras}. The masses in this case are referring to $R_{200}$. The dynamical time is defined as $t_d = \sqrt{(3/4\pi) (1/200G\rho_{\rm crit})}$, the critical density $\rho_{\rm crit}$ depends only on the cosmology, so the dynamical time does not depend on the cluster features, but evolves only with redshift. As in \cite{contreras}, the merger event can be described introducing 4 characteristic redshifts: $z_{\rm before}$, the moment right before the merger, when the main object is still in equilibrium; $z_{\rm start}$, the cluster starts growing in mass and the merger begins; $z_{\rm end}$, the merger disturbing effects end; $z_{\rm after}$, the end of the whole merger phase, when the cluster returns to equilibrium. We call merger phase the time between $z_{before} - z_{after}$ and merger event between $z_{start} - z_{end}$.

\begin{figure}
  \begin{minipage}[c]{0.55\textwidth}
\includegraphics[width=\textwidth]{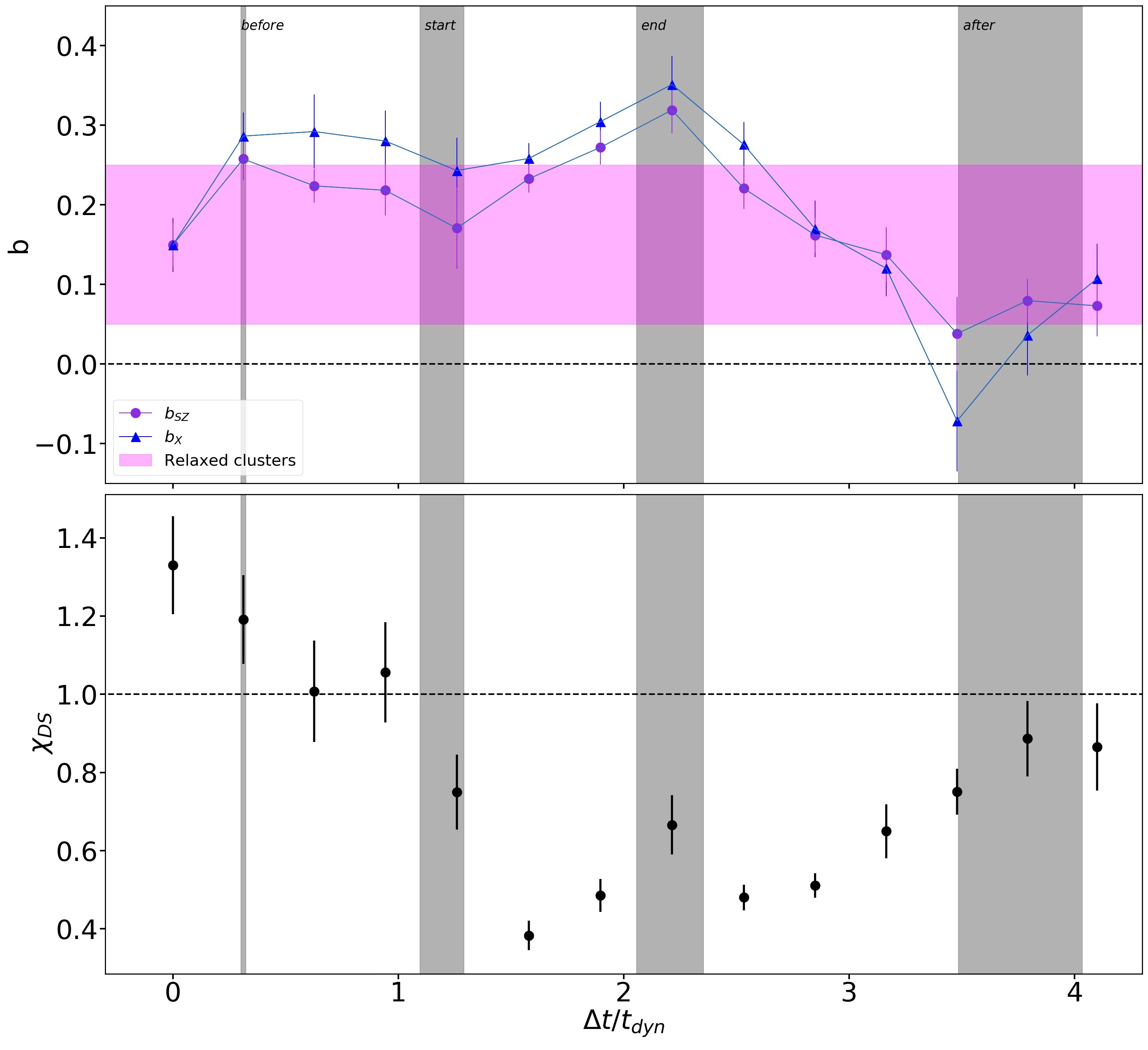}  
  \end{minipage}\hfill
  \begin{minipage}[c]{0.42\textwidth}
    \caption{Top panel: the stacked biases of 12 events at $R_{200}$ (medians and standard deviations) are represented as a function of the difference in times from $t_{\rm before-1}$, divided by the dynamical time. The SZ bias is represented in purple, while the X one in blue. The magenta shaded region shows the bias ranges for the relaxed clusters at $R_{200}$. Central panel: the stacked relaxation parameter estimated at $R_{200}$, the means and the standard deviations are represented as a function of $\Delta t /t_{dyn}$. The black shaded areas represent the $1\sigma$ regions of the averaged time before, after, start and end of the merger. } 
  \label{fig-4}
  \end{minipage}
\end{figure}

In our sample 12 clusters experience a major-merger. In the top panel of Fig. \ref{fig-4} the stacking of the two biases is represented as a function of $\Delta t / t_{\rm dyn}$, which is defined as $\Delta t / t_{\rm dyn} = (t_{(\rm before-1)+\textit{i}} - t_{(\rm before-1)}) / t_{\rm dyn}$, where $t_{\rm dyn}$ is the dynamical time and $t_{\rm before-1}$ corresponds to the analysed redshift right before $z_{before}$, namely the time before the whole merger phase. 

Before the merger phase (first point in Fig. \ref{fig-4}) the bias is in agreement with the typical relaxed clusters values. Right after $z_{before}$, the biases increase until $\sim$0.3 and stay almost constant till $z_{\rm end}$. After the end of the merger event, where the secondary object is completely within the $R_{200}$ of the main object, the relaxation process begins and the bias starts to quickly decrease, until $z_{\rm after}$ where a minimum is reached, with negative values. Here the profiles of the thermodynamic quantities (see Section \ref{subsec-5}) become steeper, due to the internal moving substructure, leading to an increase in the HE mass (see Eqs \ref{eq:Mhe_P},\ref{eq:Mhe_T}) and a decrease in the bias. After the end of the merger phase, the bias value is again compatible with the one of the relaxed clusters. This behaviour can be explained also looking at the stacked profile of the relaxation parameter, Fig.\ref{fig-4} - bottom panel. The value of $\chi_{\rm DS}$ at $z_{\rm before}$ is close to 1, pointing to a general relaxed state. At $z_{\rm start}$, it starts to decrease, so the clusters become disturbed and after $z_{\rm end}$ it increase again, even if not reaching again the original relaxation state, as \cite{contreras}.

\section{Conclusions}
\label{sec-9}
In this work the 3D profiles of temperature, pressure and gas density of a set of almost 300 cluster are analysed. These clusters are extracted from {\sc The Three Hundred} simulations. From the profiles, we recover the HE masses and their biases. The bias dependencies on redshift, dynamical state and mass are studied. We do not find correlation between the bias and these quantities, in agreement with other simulations. We also find that the very disturbed systems tend to have a negative bias, namely an overestimation of the mass with HE.  

Moreover, we analysed the bias in 12 major-mergers. Before the merger, the objects are on average relaxed. During the merger, the main object gets more and more disturbed, due to the secondary object which is being absorbed, translating into a steepening in the thermodynamic profiles. This causes an overestimation of the mass with HE, i.e. the bias reaches negative values. As soon as the merger phase ends, the cluster gets back to a more relaxed state, and the bias assumes the typical values of relaxed clusters.

%%%%%%%%%%%%%%%%%%%%%%%%%%%%%%%%%%%%%%%%%%%%%%%%%%%%%%%%%%%%%%%

\end{document}